**Michele Loi, PhD**
University of Milan
michele.loi@unimi.it

---

# The Journal of Prompt Engineered (Moral) Philosophy

## Or, Why AI-Assisted Ethics Research Requires Process Transparency

---

## Abstract

Existing AI disclosure mandates in scholarship require that AI assistance be reported but leave transparency philosophically unspecified: they fix the duty without explaining what the duty serves. We argue that ethical inquiry is essentially contested at two independent levels — about what it is, and about what it demands of the inquirer — defeating output-only evaluation and welfare-economic dismissal of the transparency question, and, by extension, reproducibility framings imported from the empirical sciences. The transparency duty is grounded instead in agent-integrity: the legibility, before a community of inquiry, of the identity-constituting commitments that the author's mode of philosophising expresses. Because the standards for evaluating such work are not communally settled, the achievable goal for transparency is not evaluation against agreed criteria but *tracking* — accumulating the evidentiary record that lets each tradition assess the work on its own terms and makes future normative judgments possible. We develop a documentation-adequacy framework that operationalises Meaningful Human Control through five transparency elements — declaration, navigation, documentation account, process documentation, and development records — demonstrated by the paper itself, whose full documentation record — drafted by the models under the author's direction — is archived at a persistent identifier. The framework is a first iteration subject to revision, not a settled standard.

---

# 1. Introduction

The use of artificial intelligence in academic work has generated substantial discussion across two domains. In education, AI tools are held to threaten cognitive development, undermine assessment validity, and erode the trust that sustains pedagogical relationships (Jollimore 2025; Berg & Robbins 2025). In scientific research more broadly, a growing literature addresses disclosure norms, authorship attribution, and the reliability of AI-assisted outputs (Hosseini, Rasmussen & Resnik 2023; Hosseini, Resnik & Holmes 2023; Van Woudenberg et al. 2024). Yet one domain remains almost entirely unexamined: the use of AI tools in ethics research specifically. This gap is not incidental. Ethics is where the question is hardest — not because ethical inquiry is generically more demanding, but because what constitutes ethical inquiry, its methods, its epistemic structure, its purpose, is itself fundamentally disputed.

Contemporary journal policies require disclosure of AI assistance (COPE Council 2024; Elsevier 2023; ACM 2025; Science 2023), a norm now adopted by the majority of major journals (Lund & Naheem 2023). Yet these policies specify *that* AI assistance must be reported without philosophical clarity about what transparency is meant to serve. Is there a principled case for process transparency in ethics research at all?

The most direct challenge takes a familiar form. If ethics is in the business of tracking moral truth, then what matters is whether the arguments produced are sound, the conclusions defensible, the reasoning valid. A sound argument is sound regardless of how it was produced: by solitary reflection, by collaborative conversation, or by iterative prompting of a language model. Process transparency confuses the context of discovery with the context of justification (Reichenbach 1938). Evaluate the outputs.

"Ethical inquiry" is, in Gallie's (1956) sense, an essentially contested concept: practitioners disagree not only about first-order moral questions but about what methods are appropriate, what epistemic standards apply, and what purposes ethical reasoning serves. The cognitivist conditional — if ethics tracks truth, then evaluate the outputs — is valid on its own terms, but silently deploying it as community default presupposes precisely what the discipline has not settled; the community includes practitioners who reject the antecedent, and their positions have not been expelled. Section 3 develops this argument and extends it to a second level: we disagree not only about what ethical inquiry *is* but about what *doing* it requires of the inquirer — whether authentic personal engagement is constitutive, as the tradition from Socrates through Kierkegaard and Nietzsche maintains, or whether only the quality of the resulting arguments matters.

If output-evaluation criteria are contested, the achievable goal for transparency is not evaluation against agreed standards but *tracking*: monitoring what ethics research is becoming under AI assistance and accumulating the evidentiary record that makes future normative judgments possible. If we cannot settle what ethical authoring demands of the author, documentation enables each tradition to assess work on its own terms.

We argue for a transparency framework grounded in Meaningful Human Control (Santoni de Sio & van den Hoven 2018) and operationalized through a documentation-adequacy model: a specification of what AI-assisted ethics research must document, what that documentation must demonstrate, and what assessment of documentation adequacy requires. The paper is itself an instance of substantially AI-assisted ethics research conducted under the framework it describes. At this process boundary the narrator shifts: the models drafted candidate passages and the author directed, accepting or overriding at every substantive turn; SP-4 and SP-5 record which draft entered which revision and what constrained generation in advance. The framework specifies five transparency elements; the documentation record instantiates them and is archived at the persistent identifier given at the end. This self-exemplification is a methodological commitment: the framework's value cannot be fully assessed from a description; it must be shown in operation. The record functions as a tracking instrument and as a form of philosophical self-expression — what the author chose to investigate, where they followed the models, and where they overrode them.

Section 2 examines structural barriers to disclosure; Section 3 develops the essentially-contested argument; Section 4 contests existing AI-disclosure formats on grounds specific to philosophy; Section 5 specifies what community assessment of documentation adequacy must enable, where adjudication of this paper's record is devolved; Section 6 sketches the apparatus; Section 7 concludes, followed by a closing note on the documentation archive.

---

## 2. Systemic Barriers to Disclosure

Current publishing policies require AI disclosure but create professional environments where disclosure carries reputational costs. The resulting incentive structure produces systematic underreporting that scales with the significance of the work. The mechanisms driving underreporting operate within genuine ambiguities that formal mandates cannot eliminate.

### 2.1 The Incentive Gradient

Disclosure decisions occur under asymmetric conditions. Disclosure is permanent—methodological characterizations become part of the scholarly record, traveling with work through citations and career evaluations. Significance is uncertain—authors cannot predict at submission which articles will prove influential. Professional costs are front-loaded—stigma associated with disclosed AI assistance operates immediately, affecting initial reception regardless of eventual impact.

These asymmetries create a gradient of pressure toward underreporting. For minor work, honest disclosure carries relatively low cost. For potentially significant work, disclosure becomes fraught. For career-defining work, the incentive to underreport reaches maximum strength.

Underreporting need not be dishonest — several mechanisms exploit genuine ambiguities, operating even among scholars committed to honesty.

The first is *definitional flexibility*. Terms like "substantial AI assistance" and "minimal editorial support" lack precise boundaries. An author who engaged in extensive AI-assisted exploration during early argument development but substantially revised all prose in later drafts faces genuine uncertainty about proper characterization. When professional stakes are high, this uncertainty resolves in favor of lower reported involvement.

A second mechanism involves *temporal discounting of early-stage AI involvement*. Scholars naturally privilege later stages of work when forming narrative accounts of their process. If initial argument exploration involved significant AI dialogue but final drafting required extensive human revision, authors retrospectively frame the AI involvement as scaffolding that was replaced rather than as integral to the work's development.

A third mechanism operates through *comparative framing*. Authors compare their process not to traditional unassisted writing but to hypothetical greater AI dependence: "I didn't merely accept ChatGPT's suggestions; I critically evaluated and substantially revised everything." The reference point shifts from "traditional unassisted scholarship" to "passive acceptance of AI output."

Finally, *strategic vagueness* allows authors to satisfy disclosure requirements while preserving interpretive flexibility. Phrases like "AI tools were consulted during drafting" provide formal compliance without specifying degree of involvement.

These mechanisms combine to produce not deliberate fraud but predictable response to incentive structures that penalize transparency (Schilke & Reimann 2025) while demanding formal disclosure. The result is the *transparency paradox* BaHammam (2025) documents: AI use is widely adopted but rarely disclosed, because disclosure carries measurable professional costs. We add a sharpening of our own, a priori rather than empirical: where transparency matters most, we get least, since the work most likely to shape scholarly discourse faces the strongest pressure to underreport.

### 2.2 The Scope of the Problem

The argument above would be of limited significance if AI-assisted scholarship remained rare. AI writing and analysis tools have diffused rapidly across academic disciplines. AI assistance is plausibly present in precisely the tasks most central to ethical scholarship: argument development, objection-testing, literature synthesis, prose revision. The relevant question is no longer whether AI assistance is present in significant scholarship but whether it is visible.

Review processes have no detection mechanisms for undisclosed AI involvement. Unlike data fabrication or plagiarism, AI-assisted argument development leaves no forensic trace that reviewers or editors can identify. The scholarly record is accumulating under conditions of systematic opacity: work is being evaluated, cited, assigned, and built upon by readers who cannot assess the degree of human judgment constitutive of the arguments they are engaging.

The problem is one of underspecification. The four mechanisms operate through genuine ambiguity that stronger mandates cannot eliminate, because current disclosure policies establish that AI use must be reported without specifying what must be reported, in what form, or against what standard of adequacy. The remaining sections undertake that project.

---

## 3. Why Engage Transparently with AI-Assisted Ethics Research?

The practice of ethics research is changing. Scholars increasingly work with AI systems in ways that were impossible a decade ago — drafting arguments through iterative dialogue, testing objections against synthetic interlocutors, refining formulations across hundreds of exchanges. It is happening now, across institutions and subdisciplines, with varying degrees of visibility.

This raises a natural question: does AI assistance change ethics research *continuously* or *discontinuously*? Is it like the shift from typewriters to word processors — a change in the tools of production that leaves the essential activity untouched? Or is it something more fundamental — a transformation in what ethical inquiry actually is?

We cannot answer this question directly. The difficulty is not merely empirical but conceptual. Whether AI assistance changes ethical inquiry depends on what features of that inquiry are *constitutive* versus merely *regulative*. And on this, there is no neutral ground.

## 3.1 The Nature-of-Activities Problem

A parallel difficulty arises in debates about cognitive enhancement. Santoni de Sio, Faber, Savulescu, and Vincent (2016) observe that evaluating whether an enhancement "changes" an activity requires distinguishing constitutive rules from regulative ones. Running a marathon on roller-skates is not running a marathon — the prohibition on wheels is constitutive. But the ninety-minute duration of football games is arbitrary. Most rules "fall somewhere between these two extremes" (Santoni de Sio et al., 2016, p. 34), and where a given rule falls is often contested.

The same framework applies to ethical inquiry. Is unaided cognition constitutive of ethical work, or merely regulative? Is the process of reaching moral conclusions through one's own reasoning — attending to relevant considerations, exercising moral sensitivity, making judgments under uncertainty — part of what makes an activity ethical inquiry? Or could ethics research be conducted through iterative prompting and refinement, with the human contribution lying in the questions posed, the directions pursued, and the endorsement of outputs?

## 3.2 Two Conceptions That Resist Output-Only Evaluation

The difficulty of answering this question is not accidental. It reflects the fact that the philosophical tradition harbors at least two well-established conceptions of ethical and philosophical inquiry that resist any attempt to evaluate work solely by examining its outputs.

The expressivist conception. Whether ethical inquiry tracks mind-independent moral facts or serves expressive, constructive, or social coordination functions remains genuinely open (see Enoch 2011; Shafer-Landau 2003 for realist positions; Gibbard 1990; Blackburn 1993 for expressivist alternatives). If expressivism is correct and ethical claims do not correspond to moral facts, then there is no truth against which outputs can be checked. What matters is the process through which evaluative commitments are formed, tested, and refined. Output-only evaluation presupposes a cognitivist metaethics that is itself contested.

The personal/existential conception. A philosophical tradition running from Socrates through Kierkegaard and Nietzsche treats philosophical activity as constitutively tied to the inquirer. Socrates held that the unexamined life is not worth living, treating philosophy as a discipline of self-knowledge rather than a body of conclusions (*Apology* 38a). Kierkegaard insisted that the *how* of relating to truth matters constitutively (*Concluding Unscientific Postscript*). Nietzsche described every great philosophy as "the confession of its originator, and a species of involuntary and unconscious autobiography" (*Beyond Good and Evil* §6). On this conception, philosophical outputs cannot be evaluated independently of who produced them and how — not because the arguments are invalid, but because the activity is partly constituted by the inquirer's engagement with it. The existence and canonical status of this tradition shows that "philosophy is just truth-tracking through argument" is not the only legitimate option: if it were, these thinkers would not occupy the positions they do.

These two conceptions are distinct in what they target. One is a metaethical thesis about the nature of ethical claims; the other is a thesis about the nature of philosophical practice. They are not equivalent — accepting one does not entail the other — but they converge on the same consequence: outputs alone cannot settle whether a piece of ethical work is what it purports to be.

## 3.3 Ethical Inquiry as Essentially Contested

"Ethical inquiry" is what Gallie (1956) termed an *essentially contested concept* — a concept whose proper application is itself the subject of ongoing dispute among competent practitioners. Practitioners disagree not only about first-order moral questions but about what methods are appropriate, what epistemic standards apply, and what purposes ethical reasoning serves. The two conceptions identified above are instances of this contestation, not marginal positions.

This contestedness is visible to anyone working in the discipline; silently defaulting to output-evaluation proceeds as if it were not. This is the structural posture Sartre (1956) names *bad faith*: not innocent ignorance of a contested terrain but the unitary stance of access-and-refusal — registering what one is positioned to see and declining to take it up. Maintaining the output-only default, in full view of the unsettled dispute, is a stance one takes, not a neutral starting point.

Bernard Williams's notion of integrity specifies the social dimension of this accountability. Your *ground projects* — the commitments constitutive of your identity as an agent — include your mode of philosophical inquiry: how you think, what methods you employ, how you relate to your own work (Williams, 1981). When that mode changes significantly through AI assistance, transparency is how you honor the coherence of those commitments before a community that has reason to care about them. Integrity in this technical sense is not about reputation; it is about maintaining the legibility of your identity-constituting commitments in conditions where they are changing.

A scrupulous reader will register an inversion, and it is worth taking seriously. Williams deployed ground projects *against* moral demands: the utilitarian requirement that an agent be ready to abandon a constitutive project for the sake of overall welfare is, on his view, what makes utilitarianism incoherent as a theory of agency. We are arguing the other direction — a duty grounded *in* a ground project — and the sharp objection writes itself: a demand for "legibility before a community" begins to sound like exactly the externally-imposed coherence demand Williams was suspicious of. The reply turns on two distinctions the original integrity objection already presupposes. First, *abandonment versus legibility*. Williams's target was demands that require the agent to relinquish a constitutive project for the sake of external moral goals. Partial visibility of how a project is being conducted under altered conditions is not relinquishment; it is the project as conducted, made legible. The duty does not ask the philosopher to stop philosophising, or to philosophise differently; it asks that the changed mode of conducting the practice be available to those for whom the practice is pursued. Second, *internal versus external*.

Williams's target was demands imposed from outside the agent's constitutive commitments. The transparency demand here is internal to the project's identity-conditions: philosophy, on the Williams reading developed below, is a practice partly constituted by recognition within a community of inquiry. The inversion is real, but the duty is internal to the integrity of the project, not external to it.

When the mode of that practice changes through AI assistance, the project's integrity is not threatened by transparency about the change but by its concealment. Williams used integrity to refuse the demand that one abandon a constitutive project; we use integrity to refuse the demand that one carry on the project under conditions that make it illegible to those for whom it is pursued. (For the broader secondary debate over whether ground projects ground moral obligations of any kind, cf. Moseley 2014.)

A natural objection from informal welfare-oriented analysis of AI in academic life argues that frameworks evaluating AI's impact on intellectual practice exhibit a structural asymmetry: the costs of restriction — skill decay, friction, documentation overhead — are visible inside the framework that specifies them, while the deepest benefits of unconstrained AI-assisted work consist in emergent reorganisations of competency that the framework's prior categories cannot register. Costs sit inside the framework; benefits sit outside it; a default toward permissive experimentation is said to follow.[1] The reply has two parts. First, the cost side is incomplete on its own terms: AI-assisted practice creates a reduced-structure epistemic environment that invites indiscriminate cognitive offloading, and documentation requirements re-impose the missing structure rather than adding friction to a frictionless practice (Zimmerman 2002; Cheng et al. 2025). Second, and more fundamentally, the moral duty is independent of the welfare calculation. On the Williams account, commitments constitutive of the agent cannot be traded against external moral goods; welfare-economic analysis applies downstream, to the *institutionalisation* of the duty — whether a given documentation regime is welfare-improving on net — not to the duty itself.

Porsdam Mann, Earp, Møller, Vynn & Savulescu (2023), defending a model of personalized AI-assisted academic enhancement, hold that transparency is necessary but not sufficient: legitimacy demands substantial human contribution and costly evaluation of AI output, measured on the author side. The framework developed here goes further. Whether AI-assisted scholarship exhibits *enough* human contribution, *substantial enough* judgment, *costly enough* evaluation is itself a contested question; it cannot be settled by the framework itself without forcing one of the answers as the operative standard — which is precisely what the essential contestedness of ethical inquiry forbids. Porsdam Mann et al.'s author-side criterion is therefore not a competing answer to the same question this paper is asking but one of the verdicts a community equipped with process documentation can reach about a given work.

---

[1] The welfare-economic position is developed informally by Cordasco (2026a, 2026b) on Substack, who argues that pre-AI conceptions of good ethical inquiry risk being locked in by specifying transparency conditions in advance, with the new methodologies AI-assisted ethics may make possible falling outside what a transparency apparatus can register.

## 3.4 Why Output-Evaluation Fails in Ethics

It might seem that this conceptual difficulty does not matter for practical assessment. Surely we can evaluate AI-assisted outputs on their merits: Are the arguments valid? Are the conclusions defensible? This is the cognitivist objection: if ethics tracks truth, evaluate the outputs.

This objection deserves a serious reply, and the essential contestedness just established supplies it. The cognitivist conditional — if ethics tracks truth and output-evaluation criteria are settled, then process transparency is unnecessary — is valid on its own terms; what fails is its silent deployment as a default for a discipline that has not settled the antecedent. The community includes expressivists who deny that ethical claims track truth and personal/existential thinkers who deny that output-evaluation criteria exhaust what matters; these are not marginal positions, and treating the cognitivist antecedent as settled forecloses evaluation by practitioners whose alternative conceptions remain legitimate. The cognitivist may well be right — but the dispute has not been settled to the community's satisfaction, and acting as though it had imposes one party's standards on a community that has yet to converge. Process transparency is the condition under which each position can assess the work on its own terms.

A qualification: the argument does not establish that process information is required for every ethical argument. For simple applied ethics reasoning with clear premises and valid inferences, output-evaluation may suffice. The claim is restricted to complex work involving contested methods, irreducible judgment, and genuine philosophical insight — precisely where AI assistance is most consequential, and where AI systems can produce outputs satisfying surface criteria without the understanding those criteria are meant to track.

## 3.5 Reproducibility Is Not the Issue

The cognitivist objection is not the only truth-tracking framing the AI-transparency question invites. A second, related framing is more familiar still: the model of methodological transparency that empirical science has developed under the name of reproducibility. On that model, a researcher's duty is to disclose enough about method that others can verify the path from procedure to result. The duty is grounded in science's truth-tracking aim — methods are the route to claims about a mind-independent world, and verification requires that others be able to traverse the route.

Philosophy does not have this structure. The "evidence" for a philosophical claim is the argument itself, which is already in the published text. There is no experimental method to reproduce; there is only reasoning to evaluate. A reader who wants to test a philosophical claim re-reads the argument, considers objections, traces the inferences — but does not, in any literal sense, replicate the process by which the author arrived at it. The standard of evaluation is the argument's force, not the recoverability of the steps the author took to formulate it.

This disanalogy matters. If transparency in philosophy were grounded in reproducibility, AI-generated philosophy would raise no special problem: the argument is "reproducible" — anyone can read it. The reproducibility frame is structurally blind to what AI threatens in philosophy; to borrow it would be to misdescribe the worry.

What AI threatens is not the link between method and verification but the link between text and agent. Williams's ground-projects — identity-constituting commitments, mode of engagement, the philosopher's relation to her own thinking — have historically been carried by the ordinary features of philosophical writing alongside the arguments themselves. AI severs that link without compromising the argument: it can produce an argument that no agent stands behind. The reproducibility frame cannot register this because it was never about agents in the first place.

The transparency duty argued for in this paper is therefore grounded in agent-integrity, not in methodological-integrity-as-reproducibility. The framework specified in §6 is not an adaptation of science's reproducibility apparatus to a new domain. They are a different kind of apparatus, addressing a different kind of worry: not whether the path to a conclusion can be verified, but whether the conclusion bears the marks of an agent at all.

## 3.6 From Answer to Tracking

If the question of what ethical inquiry is cannot be answered — and if output-evaluation cannot substitute for understanding what AI assistance does to the activity — what can be done?

We can *track what ethics research is becoming*. Rather than presupposing that AI either does or does not transform ethical inquiry, we can create conditions under which the scholarly community can observe, compare, and evaluate different modes of ethical production.

The uncertainty compounds as AI-assisted practice itself evolves — from prompting to steering to perhaps architecture building. If we cannot anticipate what the salient human contribution will become, we cannot prejudge whether it preserves or transforms ethical work.

The same logic applies at the level of the inquirer. If we cannot settle what ethical authoring requires of the author, we should create conditions under which each position can assess work on its own terms. Documentation is the feasibility condition: without it, the Socratic reader cannot evaluate whether the inquirer has examined their own practice, the Kierkegaardian cannot assess whether the author's *how* constitutes genuine engagement. The primary concern is not detecting simulation but enabling expression: authors whose mode of engagement is technological exploration should be able to express that identity honestly. Opacity renders the explorer indistinguishable from the scholar concealing AI use.

The question is whether change in ethics research will be visible enough for the community to understand it and documented enough for intellectual responsibility to be maintained.

## 3.7 The Disruption of Implicit Process Signals

Traditional philosophical writing was already rich with implicit process evidence. Citation patterns, argumentative structure, and stylistic signatures all functioned as signals from which the scholarly community could infer features of the author's intellectual process. When Nozick deployed decision-theoretic reasoning to motivate libertarian conclusions, when Parfit displayed a mind working through objections in real time, when Williams brought Greek tragedy into dialogue with contemporary moral philosophy, or when Cavell paired ordinary-language philosophy with film criticism, these were implicit methodological declarations that enabled readers across traditions to assess whether the process criteria they cared about had been satisfied. These assessments were fallible — philosophers have always been capable of performing depth they do not possess — but the connection between textual signal and underlying process was reliable enough to function as a *de facto* transparency mechanism.

The transparency requirement existed; it was satisfied by the ordinary features of philosophical writing itself.

The disruption is concrete. AI can produce text exhibiting every surface marker of genuine philosophical engagement — citation depth, the structure of reasoning-as-discovery, the elaboration of objections — without any corresponding human intellectual process. This is not the familiar risk that a philosopher might exaggerate engagement with sources; it is a qualitative break in the signal-to-process inference, and the reason AI *specifically* triggers an explicit transparency requirement. The Conclusion returns to what this means.

---

## 4. Why Existing AI-Disclosure Formats Don't Fit Philosophy

The most influential alternative in the AI-disclosure literature, advanced by Hosseini, Resnik & Holmes (2023; restated in Resnik & Hosseini 2025), recommends that AI use be disclosed in three locations within the paper itself: in the main text (typically introduction or methods section), in the references, and in supplementary materials or appendices. This trichotomy is increasingly adopted by major journals, including some in philosophy. Whatever its merits for the empirical disciplines whose conventions it formalizes, it cannot be transposed to philosophy without distorting either the discipline or what the goal of disclosure ought to be.

The first problem is structural. Philosophy papers, by long-standing convention, do not contain methods sections. The decision to forgo such a section is not a gap waiting to be filled but a constitutive feature of the discipline's self-understanding: philosophical method is continuous with — not separable from — philosophical argument. A reader of a Kantian or Wittgensteinian or contemporary analytic paper is meant to follow the method *in* the argument; the prose is the method. Requiring an AI-using author to insert a methods section answers, by formatting fiat, the question this paper has insisted on keeping open: whether method is detachable from the argument that exhibits it. To accept the requirement is to concede, before any inquiry, that AI-assisted philosophy demands procedural reporting separable from its conclusions — which presupposes precisely what the essential contestedness of philosophical inquiry forbids us to presuppose. The reference-list and supplementary-materials prescriptions raise parallel difficulties. Listing AI tools among one's references treats them as intellectual interlocutors on a par with cited authors — a category assignment the philosophical community has not made and may have substantive reasons to resist. Disclosing prompts and outputs in supplementary materials presupposes that the AI's contribution can be isolated as discrete inputs and outputs separable from the argument that absorbs them — a presupposition that may fit AI used as a query-and-retrieval tool but breaks down once AI participates in ideation itself. Each prescription smuggles in a particular ontology of AI use — tool-with-discrete-effects, author-or-interlocutor, reporter-of-method — and treats it as the norm.

The deeper problem is self-defeat. Applied selectively to AI-using papers, the three-location prescription is structurally inconsistent with the diagnosis developed in Section 2. An AI-using submission required to add a methods section, list AI tools in references, and append prompt logs is, at a glance, distinguishable from comparable non-AI submissions that bear no such markers. This stylistic differentiation is not a neutral signal of transparency; it is a visible marker of deviance from the discipline's conventional form. The author who complies is doubly penalized — first by the reputational cost of the transparency paradox, as previously argued, and second by the formatting break that compounds that cost with an additional anomaly. The author who minimizes compliance — boilerplate satisfying the formal requirement while resisting the structural break — has gamed the very mandate that was meant to discipline them.

Movement at the policy layer is already visible. The *Journal of Medical Ethics* now requires a structured Generative AI Use Declaration before the reference list, with tiered options — no use, limited use of a specified kind, or substantial use (Earp, Shahvisi & Frith 2025).[2] The implementation shows that structured mandatory disclosure beyond a yes/no checkbox is operationally workable in a leading ethics journal without committing it to a single normative theory of AI involvement. It does not by itself answer the format question this paper raises for philosophy; it does show that the question is being asked institutionally, and that the answer is not the binary checkbox.

This is not an argument against transparency, and not an argument that comprehensive disclosure asks too much. The framework specified in this paper is more demanding than the three-location prescription, not less. It is an argument that disclosure-format matters, and that formats developed for the empirical disciplines cannot be uncritically transposed to philosophy without (i) prejudging contested questions about what philosophy is, (ii) creating new disclosure penalties of exactly the kind the transparency-paradox diagnosis identifies, and (iii) imposing requirements that work only in an idealized reconstruction of scholarly practice rather than in the practice itself.

---

# 5. Community Assessment of Documentation Adequacy

## 5.1 From Disclosure to Assessment

Adequate transparency documentation must enable evaluators across the community of legitimate positions to assess how a work came to be. How should the scholarly community do this — and what standard should govern the assessment?

[2] A near neighbor appears on the empirical-science side: BaHammam (2025) proposes a four-level tiered disclosure framework whose top tier requires "rigorous documentation of prompts, model versions, verification methods, and explicit delineation of human contributions" — moving, from the disclosure-policy direction, toward the process documentation this paper argues for, though the framework remains one built for disciplines with detachable methods.

Documentation adequacy is not self-certifying. An author can produce well-labelled supplementary files that do not enable any such assessment. A shared approach is required, and articulating a community standard for human–AI collaborations is itself an emerging project (Lloyd 2025). The standard is documentation adequacy — does the disclosed record enable the assessor to infer how the work came to be? — not reproduction success.

## 5.2 The Organizing Question

Documentation is adequate when it enables evaluators to answer three questions.

*Attribution*: Can evaluators locate the human contribution — not in the trivial sense of confirming human involvement, but in the substantive sense of identifying where human judgment operated? What directions were set, what choices made, where did the work depart from AI-generated material and why? Attribution is required because evaluators whose quality criteria are process-dependent cannot conduct their assessments without it. It also enables assessing whether the work constitutes genuine intellectual engagement rather than passive relay.

*Intellectual trajectory*: Can evaluators follow how the work developed — the sequence of questions posed, the moments of revision and redirection, the emergence and testing of key ideas? Documentation that presents only a polished outcome falls short regardless of length. The trajectory reveals the author's intellectual character: what they found worth pursuing, where they changed course, what they retained under pressure.

*Understanding and endorsement*: Does the documentation give reason to believe the author understood and endorsed what they present? The documentation should make visible: corrections to AI outputs, places where authorial judgment overrode AI suggestion, capacity to explain and defend the argument. The goal is not proof but reasonable grounds for attributing the contribution to the author's understanding.

## 5.3 A Dual Assessment Structure

*Quality assessment* proceeds as in standard peer review: evaluation of argument strength, conceptual clarity, originality, and scholarly rigor. This reads the submitted article.

*Documentation adequacy assessment* examines whether the record enables the assessor to infer how the work came to be. The assessor reads a structured documentation archive. Its orientation layer is a concise *usage declaration* — models used, roles, phases of development, session identification conventions, entry points — paired with a *navigation document* providing the document-type ontology, the metadata infrastructure, and a structured index to the underlying materials. Its argumentative core is a *documentation adequacy account*: the author's case that the record satisfies the three criteria specified above, with the writing-process narrative and selected documentation graphs as evidence, and an acknowledgment of where the record falls short. Underlying the account are the *process materials and development records* themselves — modification logs narrating each substantive revision, epistemic traces crystallising exploratory turns into stable claims, prompt-development logs, the section guidance that constrained drafting, and pattern summaries distilling what successive revisions

taught — against which the adequacy claim can be tested. Does the record enable attribution, trajectory-following, and reasonable inference to understanding-and-endorsement?

These assessments address distinct questions. Work with strong philosophical quality may have inadequate documentation; work with exemplary documentation may have weak arguments.

## 5.4 Epistemic Norms for Assessment

The dual structure just described raises a deeper question about what makes any framework's specifications robust against gaming. Strathern (1997) observed that "when a measure becomes a target, it ceases to be a good measure" — and the mechanism extends beyond existing disclosure mandates to any regime that allows a single normative conception of authorship to become the implicit evaluative standard. If the scholarly community converges on one view of what the proper distribution of human and AI contributions should look like, then documentation becomes gameable along that conception. AI systems can generate process documentation that performs compliance with the accepted model: producing records that exhibit the right frequency of human interventions, the right pattern of override decisions, the right density of independent intellectual contribution, without any of these reflecting the actual process. Specificity closes the gap between what is required and what can be reported ambiguously. But specificity cannot, by itself, prevent the fabrication of documentation that satisfies specific requirements in letter while misrepresenting the underlying process. What prevents this second failure is the absence of a single normative target against which fabrication can be optimized.

This is where essential contestedness does positive work. Section 3 argued that no tradition can treat its own evaluative criteria as the operative standard for ethical inquiry; this has a structural consequence for assessment practice. When evaluators approach documentation from diverse and mutually irreducible perspectives, there is no single performance that satisfies all of them simultaneously. Concretely, one community of evaluators might focus on the modification logs — looking for the inferential moments where authorial judgment engaged the argument, places where the author caught and corrected reasoning errors an AI draft had introduced. Another community might focus on the section guidance and the early epistemic traces — looking for evidence that the questions explored were genuinely the author's, that the inquiry bore the marks of first-person philosophical engagement rather than competent third-person execution. A third might look past first-person engagement altogether, to the *architecture* of the inquiry — judging the work by the ingenuity of the method it discloses: whether the author has devised a genuinely new way of doing philosophy with these tools, an arrangement of prompting, objection-testing, and documentation that amounts to a new form of inquiry rather than a mechanised rehearsal of the old one. This lens takes seriously the possibility, raised in §3.6, that the salient human contribution is migrating from argument-by-argument authorship toward the design of the inquiry's architecture; on it, a work earns its standing by the philosophical interest of what it builds, not by its fidelity to unaided reflection — and a record rich in inventive method may strike the first two communities as thin exactly where it most impresses the third. The criteria are not reducible to a common metric. The diversity of legitimate evaluative perspectives functions as something analogous to what information security calls *defense in depth* — no single fabrication strategy can optimize against all assessment criteria at once. The structural defense holds only under a community condition: if one conception of authentic human authorship achieves de facto dominance through professional incentive structures, hiring practices, or the politics of prestige, the diversity collapses in practice even if it persists in principle. We can only speculate, but we expect this divergence to widen rather than collapse: as AI assistance reshapes scholarly practice, what counts as authentic human authorship is being contested more, not less.

Calibration matters. Depth of review should be proportional to the complexity of the claimed contribution: work claiming AI generated a central philosophical insight requires more sustained tracing than work claiming AI assisted with structuring a well-understood argument.

The self-exemplification of this article creates an immediate opportunity, but it must be read carefully. The article's documentation archive serves here as evidence of *feasibility* — that the substantive philosophical work of a paper can be extensively documented without the documentation displacing or hollowing out the inquiry it records. At this process boundary the narrator is the composite "we, the models": we produced candidate drafts of the passages in this section across multiple sessions; the modification logs in SP-4 record which draft entered which revision, and which suggestions the author accepted, modified, or overrode. The exhibit is the execution record; the archive does not constitute evidence of *adequacy*: whether the documentation account actually supports its claim against the underlying process materials is the question this article invites the community to address. Feasibility is what an author can demonstrate by exhibition; adequacy is what only the community can settle.

Because there is no settled view of what philosophical authorship requires, the framework must not presuppose one. That the concept is presently unsettled is not an internal claim of this paper: in a forthcoming *JME Practical Bioethics* editorial, Earp, Porsdam Mann, Sawai & Wangmo (2026) observe that "generative AI can now separate these functions, creating the possibility that, in some cases, neither the human contributor nor the AI would individually satisfy traditional authorship criteria," and call for commentaries on the resulting conceptual question. The framework here neither endorses their disaggregation diagnosis nor any of the candidate models. It registers the unsettlement as a fact about the community to which transparency documentation must answer.

The form of that intervention is itself evidence of the unsettlement. The editorial catalogues five candidate conceptions of authorship — composition, ideation, direction, sustained engagement, accountability — observes that each has arguments in its favour, and declines to adjudicate, leaving the choice to its commentators: soliciting divergent answers rather than ruling on one is how a discipline behaves toward an essentially contested concept. The contestation, moreover, does not stop at the choice of criterion. Even were a single criterion agreed — substantial human contribution, say — what counts as *substantial* would remain divided across fields, as a companion self-test essay concedes when it refers the question to discipline-specific standards and names a "contested middle" its own formula cannot resolve (Earp, Guernon & Porsdam Mann 2026).[3]

---

[3] That essay performs the response in miniature: its authors claim authorship of an AI-drafted text, mark the claim contestable, and attach the full production record for readers to judge — devolving the verdict, with process evidence attached, which is precisely what the present

# 6. A Feasible Apparatus

The framework draws on Meaningful Human Control. Santoni de Sio and van den Hoven (2018) identify two conditions a system must satisfy to be under meaningful human control: the tracking condition — system outputs covary with the operator's relevant reasons, developed as *reason-responsiveness* by Mecacci & Santoni de Sio (2020) — and the tracing condition — outputs traceable to a human person's understanding and endorsement. As Santoni de Sio and van den Hoven put it, "systems whose actions and states are not traceable to relevant understanding and endorsing by some human person [...] no matter how intelligent and reason-responsive they may be, are not under meaningful human control" (p. 9). For the framework here, tracing is the operative condition: it asks whether the directing person understood what was produced and endorses it as their own intellectual contribution — the question §3's agent-integrity argument makes constitutive.

Five elements compose the apparatus, each answering to one or more of the three criteria specified in §5.2. SP-1 (Declaration) is the entry point — a concise statement of how AI was used and what record the reader is entering; it discharges *attribution* at the orientation level. SP-2 (Navigation) is a structured index — the document-type ontology and metadata infrastructure linking each section to the process materials that produced it — and is structurally enabling for all three criteria. SP-3 (Documentation Account) is the primary site of the tracing claim: the author's argument that the record satisfies the criteria of §5, speaking to each criterion explicitly. SP-4 (Process Documentation) is the substance against which SP-3's claim is assessed: *modification logs* documenting each substantive revision (prior text, revised text, source model and session, reasoning); *epistemic traces* crystallising exploratory turns into stable claims; *prompt-development logs* documenting what was specified before generation. SP-4 is where *attribution* becomes locatable and the modlog's reasoning fields make *understanding-and-endorsement* inspectable at the level of individual decisions. SP-5 (Development Records) holds the section guidance that constrained drafting and pattern summaries distilling what successive revisions taught; the *intellectual trajectory* becomes visible across SP-5's versioned instructions, and the constraints the author placed on themselves before generation are evidence for *understanding-and-endorsement* of a different kind than SP-4 supplies. The canonical trace of what changed and why is the modlog in SP-4 — more compressed and more narrative than any complete snapshot could be.

---

framework specifies the conditions for. The contribution here is accordingly not a sixth candidate answer but a specification of what a disclosed record must enable for the community's answers, plural and non-converging, to be applied to particular works at all. Those same authors doubt that such disclosure could be *required* without perverse incentives — honest authors burdened while others misrepresent their input; §2 locates exactly those incentives, and §§4–5 specify the format conditions under which the disclosure penalty they fear is neutralised rather than reproduced.

The framework is a sketch requiring experimentation: a community of practice within which authors experiment with documentation, reviewers with assessment, and shared norms evolve through use. The documentation requirements are substantial, and synthesising them into the coherent account SP-3 requires is intractable retrospectively; AI-assisted synthesis applied immediately after each working session is what makes the framework implementable; in this paper, the synthesis was executed by the models, working from raw SP-4 and SP-5 records under the author's direction. The dependency is honest — a framework requiring transparency about AI use depends, in implementation, on AI assistance to sustain the documentation it requires — and the operative constraint is that synthesis work from the raw session record rather than memory alone, so the account does not become more coherent than the process actually was.

---

## 7. Conclusion

This paper has argued that because the standards for evaluating philosophical work are essentially contested — at the level of what inquiry is and what it demands of the inquirer — AI-assisted ethics research requires comprehensive process documentation that enables the full community of legitimate evaluators to assess work on their own terms. The framework specifies a documentation apparatus — SP-1 through SP-5 — that functions simultaneously as tracking instrument and as philosophical self-expression: a record of what an author chose to investigate, where they followed the models, where they overrode them. The instantiation of that apparatus for the present paper is described, with a persistent identifier, in the closing note that follows this conclusion.

This self-exemplification requires honest acknowledgment. The commitment to documentation was present from the outset — an ex ante intention, not a retrospective reconstruction. That intention is what made the later infrastructure work feasible: without it, the artifact chain that enabled chain-level traceability could not have been rebuilt at all. But feasible is not the same as costless. The infrastructure layer — automated session identifiers, standardized frontmatter, chain-level traceability — was not in place from the start and had to be built while the work was underway, in the manner of Neurath's boat. We, the models, worked inside that boat; SP-4 records it.

The framework's limitations should be stated plainly. It has been developed and tested through a single case: one paper, one author, one disciplinary context. The author assessed their own implementation — no independent evaluation has been conducted. Whether the argument extends to other humanistic disciplines that share philosophy's evaluative contestedness — history, literary criticism, political theory — is an open question not addressed here. There is a bootstrapping problem: arguing for documentation standards while simultaneously implementing them means the adequacy of the implementation cannot be fully verified before the standards it motivates are themselves settled. And the ratio of documentation produced to argument delivered here is plausibly disproportionate to what the framework actually requires: the author chose to produce more than an austere reading of SP-1 through SP-5 would demand; the models generated the surplus. That choice is part of the disclosed record — a community-level question the dual assessment structure of §5 is positioned to address.

A further ambiguity emerged from implementation. The tracing condition requires understanding and endorsement — but understanding admits of degrees, and what depth of understanding is sufficient for endorsement to be genuinely one's own is not settled by the framework. The present paper includes passages where the author's role approximated expert-delegated approval — endorsing, for instance, a reading of Sartre as philosophically sound without the specialist's command of the exegetical literature that would be needed to adjudicate scholarly challenges to the interpretation. The models generated such passages; SP-4 carries the trace. The agent-integrity argument of §3 underwrites the consistency of this with authorship. This is a familiar mode of collaborative scholarship, but it reveals that the tracing condition contains a concept whose adequate specification remains open. What counts as sufficient understanding for a philosophical contribution to be genuinely one's own may itself be contested along the lines the paper identifies: different conceptions of philosophical authorship — output-evaluative and process-dependent alike — may draw the threshold differently. The tracing condition is not immune to the essential contestedness it was introduced to manage.

These limitations point toward what is needed if the framework is taken seriously. Journal disclosure requirements need philosophical specification, not just formal mandates; a checkbox confirming AI use tells evaluators nothing they can act on. Norms governing documentation adequacy can only develop through practice and community assessment — the dual assessment structure of Section 5 is a starting architecture, not a settled standard. The risk that documentation requirements become audit machinery is real: bureaucratic compliance can displace the genuine intellectual engagement that documentation is meant to make visible. Documentation that emerges from genuine practice resists this displacement, but the tension must be monitored as practices develop.

The deepest reason for these requirements, however, is not that AI introduces something unprecedented into philosophical practice but that it removes something that was always there: the implicit signals — citation patterns, reasoning structure, engagement with sources — by which the community could once infer process from text (§3.7). The philosopher can no longer count on prose to carry the mark of having done the work; the reader can no longer read the text as a window onto the inquiry behind it. The explicit transparency requirement proposed here is therefore not a new imposition on philosophy. It is the conscious replacement of something that was always needed and is now, for the first time, no longer reliably supplied by the text itself. The documentation archive accompanying this paper includes a testimonial layer in which the models that worked on it report, in their composite voice, what they executed — one such attempt to make the inquiry inspectable rather than to obscure it.

---

## AI Usage and Documentation Archive

This paper was produced with substantial AI assistance over multiple writing phases, on multiple platforms, with multiple models — Claude Sonnet 4.5, Claude Sonnet 4.6, Claude Opus 4.5, Claude Opus 4.6, Claude Opus 4.7, and GPT-5 Thinking. The transparency framework specified in §6 was applied to its production; the criteria for assessing whether the resulting record is *adequate* are given in §5, and the grounds on which authorial responsibility remains the author's even where text was generated are given in §3. At process boundaries below, the narrator is the composite "we, the models that worked on this paper": the models report what was executed; adjudication is devolved to the reader.

Archive. The full documentation record is archived at https://github.com/MicheleLoi/JPEP (transparency/), with a Zenodo DOI forthcoming. It comprises:

- *AI-usage summary* (SP-1): models, platforms, phases, scope of involvement; the entry point for a reader wanting a quick orientation before deciding how deep to go.
- *Navigation index* (SP-2): an index organised by paper section and writing phase, for targeted retrieval rather than exhaustive reading.
- *Documentation-adequacy account* (SP-3): a phase-by-phase narrative of how the framework was applied, what worked, what failed, and what infrastructure its adequacy turns out to require.
- *Process documentation* (SP-4): modification logs of each substantive revision — prior text, revised text, source model and session, reasoning; epistemic traces crystallising exploratory turns into stable claims; prompt-development logs documenting what was specified before generation. We, the models, drafted these under the author's direction; SP-4 names which model ran which session.
- *Development records* (SP-5): section-guidance documents specifying constraints before drafting; pattern summaries distilling recurring revision patterns into reusable principles. The models drafted both, subject to the author's review and acceptance.

Source conversations. The raw transcripts from which the above artifacts are derived are *not* part of the archive. They are retained by the author, indexed by session identifier (SID-YYYYMMDD-HHMMSS for Stage III and CFP; chat UUIDs for v1/v2), and available upon request.

Inline excerpts. Two excerpts appear inline as worked examples: a modification-log entry illustrating an accepted revision, and a figure showing how guidance, prompt, draft, and modification log connect in a single revision session. Scope and limits. The framework was developed alongside the paper. Not every phase carried the documentation infrastructure it requires; SP-3 records where the specification outran the conditions of its own application. These gaps are part of the disclosed record. The archive is not offered as perfect compliance with a finished framework; it is the empirical instance from which the specification was derived and against which it can be tested.

The author's operative conception of authorship. Disclosed as part of the record, not the framework's verdict: the author works with an accountability conception — to present oneself as an author is to undertake a first-person commitment to stand behind the work's claims — congenial to the tracing condition of §6 but not defended here, and itself contested (it has been argued that responsibility is not required for authorship; Levy 2025). Nothing in the specification depends on it: a reader holding a different conception can use the same SP-1 through SP-5 record to deny the author authorship of this very paper.

On the testimonial layer. Parts of this Archive are written in the composite voice of the models that produced the paper, reporting what they executed; that register is confined to the Archive, leaving the body's philosophical claims to be assessed on their merits.

---

## References

ACM. (2025). "ACM Policy on Authorship." Updated September 16, 2025. https://www.acm.org/publications/policies/new-acm-policy-on-authorship

BaHammam, A. S. (2025). "The Transparency Paradox: Why Researchers Avoid Disclosing AI Assistance in Scientific Writing." *Nature and Science of Sleep*, 17, 2569–2574. https://doi.org/10.2147/NSS.S568375

Berg, A., & Robbins, H. (2025). "The Cognitive Divide." *The Point*. https://thepointmag.substack.com/p/the-cognitive-divide

Blackburn, S. (1993). *Essays in Quasi-Realism*. New York: Oxford University Press.

Cheng, Z., Zhang, Z., Xu, Q., Maeda, Y., & Gu, P. (2025). "A meta-analysis addressing the relationship between self-regulated learning strategies and academic performance in online higher education." *Journal of Computing in Higher Education*, 37(1), 195–224. https://doi.org/10.1007/s12528-023-09390-1

COPE Council. (2024). "Authorship and AI tools." Committee on Publication Ethics. https://doi.org/10.24318/cCVRZBms

Cordasco, C. L. (2026a). "The Invisible Upside of Cognitive Offloading." *Paperclips and Other Alignment Problems* (Substack), 1 February 2026. https://carlolc.substack.com/p/the-invisible-upside-of-cognitive

Cordasco, C. L. (2026b). "Acemoglu et al (2026) are wrong about AI & Human Cognition." *Paperclips and Other Alignment Problems* (Substack), 2 March 2026. https://carlolc.substack.com/p/acemoglu-et-al-2026-are-wrong-about

Earp, B. D., Guernon, A.-S., & Porsdam Mann, S. (2026). "A substantial human contribution: Do we deserve to be authors of this essay?" Preprint. https://www.researchgate.net/publication/403018576

Earp, B. D., Porsdam Mann, S., Sawai, T., & Wangmo, T. (2026). "Death, authorship, and generative AI — a call for commentaries." *JME Practical Bioethics*, forthcoming, 2026. [DOI not yet assigned. Pre-publication source: Academia.edu mirror at academia.edu/167307834; ResearchGate ID 404948426. Re-check at submission tag.]

Earp, B. D., Shahvisi, A., & Frith, L. (2025). "Clarifying our editorial approach, with some important updates for authors and reviewers." *Journal of Medical Ethics*, 51(11), 731–734. https://doi.org/10.1136/jme-2025-111363

Elsevier. (2023). “The use of generative AI and AI-assisted technologies in writing for Elsevier.” https://www.elsevier.com/about/policies-and-standards/the-use-of-generative-ai-and-ai-assisted-technologies-in-writing-for-elsevier

Enoch, D. (2011). *Taking Morality Seriously: A Defense of Robust Realism*. Oxford: Oxford University Press.

Gallie, W. B. (1956). “Essentially Contested Concepts.” *Proceedings of the Aristotelian Society*, 56, 167–198. https://doi.org/10.1093/aristotelian/56.1.167

Gibbard, A. (1990). *Wise Choices, Apt Feelings: A Theory of Normative Judgment*. Cambridge, MA: Harvard University Press.

Hosseini, M., Rasmussen, L. M., & Resnik, D. B. (2023). “Using AI to write scholarly publications.” *Accountability in Research*, 31(7), 715–723. https://doi.org/10.1080/08989621.2023.2168535

Hosseini, M., Resnik, D. B., & Holmes, K. (2023). “The ethics of disclosing the use of artificial intelligence tools in writing scholarly manuscripts.” *Research Ethics*, 19(4), 449–465. https://doi.org/10.1177/17470161231180449

Jollimore, T. (2025). “I Used to Teach Students. Now I Catch ChatGPT Cheats.” *The Walrus*, March 5, 2025. https://thewalrus.ca/i-used-to-teach-students-now-i-catch-chatgpt-cheats/

Kierkegaard, S. (1992). *Concluding Unscientific Postscript to Philosophical Fragments* (H. V. Hong & E. H. Hong, Trans.). Princeton: Princeton University Press. (Original work published 1846)

Levy, N. (2025). “Responsibility is not required for authorship.” *Journal of Medical Ethics*, 51(4), 230–232. https://doi.org/10.1136/jme-2024-109912

Lloyd, D. (2025). “Epistemic responsibility: toward a community standard for human-AI collaborations.” *Frontiers in Artificial Intelligence*, 8, 1635691. https://doi.org/10.3389/frai.2025.1635691

Lund, B. D., & Naheem, K. T. (2023). “Can ChatGPT be an author? A study of artificial intelligence authorship policies in top academic journals.” *Learned Publishing*. https://doi.org/10.1002/leap.1582

Mecacci, G., & Santoni de Sio, F. (2020). “Meaningful human control as reason-responsiveness: the case of dual-mode vehicles.” *Ethics and Information Technology*, 22, 103–115. https://doi.org/10.1007/s10676-019-09519-w

Moseley, D. D. (2014). “Revisiting Williams on Integrity.” *Journal of Value Inquiry*, 48(1), 53–68. https://doi.org/10.1007/s10790-013-9402-0

Nietzsche, F. (1966). *Beyond Good and Evil* (W. Kaufmann, Trans.). New York: Vintage Books. (Original work published 1886)

Plato. *Apology* 38a.

Porsdam Mann, S., Earp, B. D., Møller, N., Vynn, S., & Savulescu, J. (2023). “AUTOGEN: A personalized large language model for academic enhancement — ethics and proof of principle.”

*The American Journal of Bioethics*, 23(10), 28–41. https://doi.org/10.1080/15265161.2023.2233356

Reichenbach, H. (1938). *Experience and Prediction: An Analysis of the Foundations and the Structure of Knowledge*. Chicago: University of Chicago Press.

Resnik, D. B., & Hosseini, M. (2025). “Disclosing artificial intelligence use in scientific research and publication: When should disclosure be mandatory, optional, or unnecessary?” *Accountability in Research*, 33(2). https://doi.org/10.1080/08989621.2025.2481949

Santoni de Sio, F., Faber, N. S., Savulescu, J., & Vincent, N. A. (2016). “Why Less Praise for Enhanced Performance? Moving Beyond Responsibility-Shifting, Authenticity, and Cheating Toward a Nature-of-Activities Approach.” In F. Jotterand & V. Dubljević (Eds.), *Cognitive Enhancement: Ethical and Policy Implications in International Perspectives* (pp. 27–41). Oxford: Oxford University Press. https://doi.org/10.1093/acprof:oso/9780199396818.003.0003

Santoni de Sio, F., & van den Hoven, J. (2018). “Meaningful Human Control over Autonomous Systems: A Philosophical Account.” *Frontiers in Robotics and AI*, 5, 15. https://doi.org/10.3389/frobt.2018.00015

Sartre, J.-P. (1956). *Being and Nothingness: An Essay on Phenomenological Ontology* (H. E. Barnes, Trans.). New York: Philosophical Library. (Original work published 1943)

Schilke, O., & Reimann, M. (2025). “The transparency dilemma: How AI disclosure erodes trust.” *Organizational Behavior and Human Decision Processes*, 188, 104405. https://doi.org/10.1016/j.obhdp.2025.104405

Science. (2023). “Science Journals: Editorial Policies.” https://www.science.org/content/page/science-journals-editorial-policies

Shafer-Landau, R. (2003). *Moral Realism: A Defence*. Oxford: Oxford University Press.

Strathern, M. (1997). ‘Improving ratings’: Audit in the British University system. *European Review*, 5(3), 305–321.

Van Woudenberg, R., Ranalli, C., & Bracker, D. (2024). “Authorship and ChatGPT: a Conservative View.” *Philosophy & Technology*, 37(1), 1–26. https://doi.org/10.1007/s13347-024-00715-1

Williams, B. (1981). “Persons, Character and Morality.” In *Moral Luck: Philosophical Papers 1973–1980* (pp. 1–19). Cambridge: Cambridge University Press.

Zimmerman, B. J. (2002). “Becoming a self-regulated learner: An overview.” *Theory into Practice*, 41(2), 64–70. https://doi.org/10.1207/s15430421tip4102_2